\documentclass[useAMS,usenatbib]{mn2e}

\usepackage{graphicx}
\usepackage{amsmath}

\title[TeV Gamma-rays from accreting magnetars]
{TeV Gamma-rays from accreting magnetars in massive binaries}
\author[W. Bednarek]
{W. Bednarek\\
Department of Astrophysics, University of \L \'od\'z,
ul. Pomorska 149/153, 90-236 \L \'od\'z, Poland; bednar@fizwe4.phys.uni.lodz.pl}
\begin{document}

\date{Accepted . Received ; in original form }

\pagerange{\pageref{firstpage}--\pageref{lastpage}} \pubyear{2007}

\maketitle

\label{firstpage}

\begin{abstract}
We consider the neutron star (NS) of the magnetar type inside the massive binary system. We determine the conditions under which the  matter from the stellar wind can penetrate the inner magnetosphere of the magnetar. At some distance from the NS surface, the magnetic pressure can balance the gravitational pressure of the accreting matter creating very turbulent, magnetized transition region. It is suggested that this region provides good conditions for acceleration of electrons to relativistic energies.
These electrons lose energy on the synchrotron process and the Inverse Compton (IC) scattering of the radiation from the nearby massive stellar companion, producing high energy radiation from the X-rays up to $\sim$TeV $\gamma$-rays. The  primary $\gamma$-rays can be farther absorbed in the stellar radiation developing the IC $e^\pm$ pair cascade. We calculate the synchrotron X-ray emission from primary electrons and secondary $e^\pm$ pairs and the IC $\gamma$-ray emission from the cascade process.
It is shown that the quasi-simultaneous observations of the TeV $\gamma$-ray binary system
LSI +61 303 in the X-ray and the TeV $\gamma$-ray energy ranges can be explained in such an
accreting magnetar model.
\end{abstract}

\section{Introduction}

The accretion process  can occur onto the neutron star in the case of relatively slowly rotating objects even if the neutron stars are characterized by a super-strong surface magnetic field (so called magnetars). Up to now, magnetars have not been observed inside the binary systems.
However, very recently it was claimed that X-ray emission characteristic for magnetars
has been observed from the object LSI +61 303, which belongs to the rare class of massive binary systems emitting TeV $\gamma$-rays (Dubus \& Giebels~2008). The observed outburst of thermal X-ray emission might be due 
to the transient accretion of matter onto the surface of the neutron star.
Moreover, quasi-periodic X-ray emission, with the frequency of 0.2 Hz, has been reported from the direction of this binary (Ray, Hartman et al.~2008). Whether this emission is in fact related to this massive binary, it is at present an open issue (see Rea \& Torres~2008). However, if it is true, then it makes sense to speculate that the compact object inside the LSI +61 303 is an accreting magnetar.

The aim of this paper is to investigate such a hypothesis.
We apply the general scenario for the $\gamma$-ray production similar to that one recently
proposed for accreting classical and millisecond neutron stars (Bednarek 2008).
It is shown that also accreting magnetars can accelerate electrons in the magnetized, turbulent region inside the NS magnetosphere when the gravitational pressure of the accreting matter is balanced by the magnetic pressure of the rotating NS magnetosphere. We calculate the high energy radiation produced in the synchrotron and inverse Compton processes by primary electrons and secondary $e^\pm$ cascade pairs immersed in the magnetic fields of the magnetar and the massive star, and in the thermal radiation from the massive star surface.

\section{Description of the model}

We consider a compact binary system containing rotating strongly magnetized neutron star 
(magnetar) and a massive companion of the O,B type star. 
It is assumed that a mass from the 
stellar wind is effectively captured by the strong gravitational potential of the neutron star (NS). Depending on its period and surface magnetic field, the accretion process onto NS can occur in different phases. 
According to the classification scheme of Lipunov~(1992), the accretion process can occur
in the phase of accretor (for relatively slow rotators) or 
in the phase of propeller. 
The accretor phase in the case of classical and millisecond pulsars has been recently discussed by Bednarek~(2008). 
In this paper we concentrate on the propeller phase for the strongly magnetized neutron stars, i.e. magnetars.
In this case, the matter from the stellar wind can 
penetrate below the light cylinder radius of the rotating NS magnetosphere. This matter 
extracts rotational energy from the NS as a result of interaction of the free falling matter with the rigidly rotating inner NS magnetosphere. 
As a result of this interaction, a very turbulent and magnetized transition region is formed. We consider possible acceleration of electrons in such a medium. 

The distance at which the magnetic field starts to dominate the dynamics of the in-falling matter (the Alfven radius) can be estimated by comparing the magnetic field energy density with the kinetic energy density of the wind,

\begin{eqnarray}
B_{\rm A}^2/8\pi = \rho v_{\rm f}^2/2,
\label{eq1}
\end{eqnarray}
\noindent
where $B_{\rm A}$ is the magnetic field in the inner neutron star magnetosphere, 
$\rho = {\dot M}_{\rm acc}/(\pi R_{\rm A}^2v_{\rm f})$ is the density of the accreting matter, $v_{\rm f} = (2GM_{\rm NS}/R_{\rm A})^{1/2}$ is the free fall velocity of the accreting matter, $R_{\rm A}$ is the Alfven radius, and G is the gravitational constant. The matter in this transition region is very turbulent and strongly magnetized providing good conditions for acceleration of particles to high energies. Let us estimate the location of this region from the surface of the neutron star. By applying Eq. \ref{eq1}, and assuming that the magnetic field in the neutron star magnetosphere is of the dipole type, i.e. 
$B_{\rm A} = B_{\rm NS} (R_{\rm NS}/R_{\rm A})^3$, we obtain
\begin{eqnarray}
R_{\rm A} = 5.5\times 10^9 B_{\rm 14}^{4/7}M_{\rm 16}^{-2/7}~~~~{\rm cm},
\label{eq2}
\end{eqnarray}
\noindent
where the magnetic field at the neutron star surface is $B_{\rm NS} = 10^{14}B_{14}$ G 
and the accretion rate is $M = 10^{16}M_{\rm 16}$ g s$^{-1}$ (see also Baan \& Treves~1973). Then, we can estimate the magnetic field strength at the transition region,
\begin{eqnarray}
B_{\rm A} = 610M_{\rm 16}^{6/7}B_{\rm 14}^{-5/7}~~~~{\rm G}.
\label{eq3}
\end{eqnarray}
The accretion according to the propeller scenario can occur provided that some conditions are fulfilled. At first, the radius of the transition region has to 
lay inside the light cylinder radius of the neutron star, i.e $R_{\rm A} < R_{\rm LC} = cP/2\pi$, where $P = 1P_1$ s is the rotational period of the neutron star, and $c$ is the velocity of light. The above condition is fulfilled when, 
\begin{eqnarray}
P_{1} > 1.2  B_{\rm 14}^{4/7}M_{\rm 16}^{-2/7}.
\label{eq4}
\end{eqnarray}

At second, the rotational velocity of the magnetosphere at $R_{\rm A}$ has to be longer than the keplerian velocity of the accreting matter. The rotational velocity, 
\begin{eqnarray}
v_{\rm rot} = 2\pi R_{\rm A}/P\approx 3.5\times 10^{10} B_{14}^{4/7} 
M_{16}^{-2/7}/P_1~~~{\rm cm~s}^{-1}.
\label{eq5}
\end{eqnarray}
is larger than the keplerian velocity,
 \begin{eqnarray}
v_{\rm k} = (GM_{\rm NS}/R_{\rm A})^{1/2}\approx  1.8\times 10^8B_{14}^{-2/7} M_{16}^{1/7}~~~{\rm cm~s}^{-1}.
\label{eq6}
\end{eqnarray}
for the NS with the rotation periods,
\begin{eqnarray}
P_1 < 192 B_{14}^{6/7}M_{16}^{-3/7}.
\label{eq7}
\end{eqnarray}
\noindent
This last condition on the NS period separates the propeller phase (lower periods) from the accretor phase (larger periods).

The third condition relates the Alfven radius, $R_{\rm A}$, to the capturing radius of the matter from the stellar wind by the NS. It is determined by the balance between the kinetic energy of wind with its potential energy in the gravitational field of the NS, 
\begin{eqnarray}
R_{\rm c} = 2GM_{\rm NS}/v_{\rm w}^2\approx 3.7\times 10^{10}v_8^{-2}~~{\rm cm},  
\label{eq8}
\end{eqnarray}
\noindent
where $v_{\rm w} = 10^8v_8$ cm s$^{-1}$ is the velocity of the surrounding matter measured in respect to the NS. The accretion from the stellar wind occurs when $R_{\rm c} >  R_{\rm A}$, which happens for, 
\begin{eqnarray}
B_{14} < 28v_8^{-7/2}M_{16}^{1/2}.  
\label{eq9}
\end{eqnarray}
\noindent
This condition is not restrictive even for magnetars in the case of the accretion rates
considered below.

In the conditions expected for the transition region (strongly magnetized and very turbulent medium), particles should be efficiently accelerated. In this paper we consider only acceleration of electrons. The acceleration rate of electrons with energy $E$ (and the Lorentz factor $\gamma$) can be parametrized by,
\begin{eqnarray}
{\dot P}_{\rm acc} = \xi c E/r_{\rm L}
\approx 970\xi_{-1}M_{16}^{6/7}B_{14}^{-5/7} ~~~{\rm erg~s}^{-1},
\label{eq10}
\end{eqnarray}
\noindent
where $\xi = 10^{-1}\xi_{-1}$ is the acceleration parameter, $c$ the velocity of light, 
$r_{\rm L} = E/eB_{\rm A}$ the Larmor radius, and $e$ electron charge. During the acceleration process 
electrons suffer energy losses mainly on the synchrotron process and on the inverse 
Compton scattering  of radiation from the massive star.

Electrons lose energy on IC process in the Thomson (T) and the Klein-Nishina (KN) regimes.
 Let us estimate the photon energy densities from the massive star ($\rho_\star$) in the location of the NS,
\begin{eqnarray}
\rho_\star = {{4\sigma T_\star^4}\over{c}}\left({{R_\star}\over{D}}\right)^2
\approx 6.1\times 10^3T_4^4\left({{R_\star}\over{D}}\right)^2~~{\rm erg~cm}^{-3},
\label{eq11}
\end{eqnarray}
\noindent
where $T_\star = 3\times 10^4 T_4$ K and $\sigma$ is the Stefan-Boltzmann constant,
and $D$ is the distance of the NS from the massive star.

Let us also estimate the energy density of the magnetic field at the transition region (see Eq.~\ref{eq3}),
\begin{eqnarray}
\rho_{\rm B} = B^2_{\rm A}/8\pi
\approx 1.4\times 10^4 M_{16}^{12/7}B_{14}^{-10/7}~~{\rm erg~cm}^{-3},
\label{eq12}
\end{eqnarray}
\noindent
The energy losses for each process (synchrotron and IC in the T regime) can be calculated
from,
\begin{eqnarray}
{\dot P}_{\rm loss} = (4/3)c\sigma_{\rm T}\rho \gamma^2\approx 2.7\times 10^{-14}\rho_{\rm (cap, B, \star)}\gamma^2~~{\rm erg~s}^{-1},
\label{eq13}
\end{eqnarray}
\noindent
where $\sigma_{\rm T}$ is the Thomson cross section. 

The energy losses of electrons at large energies are dominated by the synchrotron process
since the IC losses in the radiation field of the massive star decline due to the Klein-Nishina cross section. Therefore, the maximum energies of accelerated electrons are determined by the balance between energy gains (Eq.~\ref{eq10}) and energy losses on synchrotron process (Eq.~\ref{eq13}). They can be expressed by, 
\begin{eqnarray}
\gamma_{\rm max}\approx 1.5\times 10^6 
\xi_{-1}^{1/2}B_{14}^{5/14}M_{16}^{-3/7}.
\label{eq14}
\end{eqnarray}
\noindent
It is clear that for the realistic parameters of the model electrons can reach TeV energies. 

The maximum power available for acceleration of electrons is limited by the energy 
extracted from the rotating neutron star by the in-falling matter. This matter from the stellar wind is accelerated to the velocity of the magnetic field lines at $R_{\rm A}$. 
The power which has to be transfered from the rotating NS to the matter can be estimated from
\begin{eqnarray}
L_{\rm acc} = {\dot M}_{\rm acc} v_{\rm rot}^2/2 \approx 6\times 10^{36}B_{14}^{8/7}M_{16}^{3/7}P_1^{-2}~~{\rm erg~s}^{-1}.
\label{eq15}
\end{eqnarray}
\noindent
By Using Eq.~\ref{eq4}, we can estimate the maximum power which can be extracted in terms of such a model in the propeller stage,
\begin{eqnarray}
L_{\rm acc}\approx 4\times 10^{36}M_{16}~~{\rm erg~s}^{-1}.
\label{eq16}
\end{eqnarray}
A part of this maximum possible power, $\eta$, can go on acceleration of electrons.

The total rotational energy of the magnetar is,
\begin{eqnarray}
P_{\rm rot} = I \omega^2/2 \approx 4.4\times 10^{46}P_1^{-2}~~{\rm erg}.
\label{eq17}
\end{eqnarray}
\noindent
where $I$ is the angular momentum of inertia of rotating NS
calculated for its mass equal to 1.4$M_\odot$ and the radius of 10 km, and 
$\omega = 2\pi/P$ is its angular velocity. This energy reservoir is enough to provide sufficient power to the accreting matter for a time
of the order of thousand of years. Note that a neutron star in a specific binary system can come through the cycles of rotational energy losses (during the propeller phase) and energy gains (during the accretor phase). These phases depend on the accretion rate from the massive star which can change due to the stellar variability or evolution. So then, the angular velocity of magnetars may also increase at some periods as in the case of millisecond pulsars.

For the known parameters of the binary system, we can estimate the accretion rate onto the magnetar, 
\begin{eqnarray}
{\dot M}_{\rm acc} = R_{\rm c}^2 {\dot M}_\star/(4D^2)\approx 5\times 10^{15} M_{-7}/(v_8
D_{12})^2~~~{\rm g~s}^{-1}.
\label{eq18}
\end{eqnarray}
\noindent
where the capturing radius is given by Eq.~\ref{eq8}, $D = 10^{12}D_{12}$ cm is the separation of the stars inside binary system, 
${\dot M}_\star = 10^{-7}{\dot M}_{-7}$ M$_\odot$ yr$^{-1}$ is the mass loss rate of the massive star, and $v_{\rm w} = 10^8v_8$ denotes the wind velocity and/or the NS velocity on its orbit around the massive star $v_8\approx 0.37D^{-1/2}_{12}$. For the luminous stars, the wind velocity, of the order of $\sim 10^3$ km s$^{-1}$, is usually larger than the NS velocity. 
The B type stars, present inside  the TeV $\gamma$-ray binaries, have typical mass loss rate 
in the range $10^{-(6-7)}$ M$_\odot$ yr$^{-1}$ and stellar wind velocities $\sim (1-3)\times 10^3$ km s$^{-1}$. Then, the typical accretion rates onto the NS are in the range 
$\sim 10^{13-17}$ g s$^{-1}$, for the separation of the companion stars below 10 radii of the massive star.

\section{X-ray and Gamma-ray emission}

In this model, radiation is produced by electrons in the synchrotron and IC process. 
We assume that primary electrons in the transition region reach the power law spectrum, $\propto E^{-s}$ up to $E_{\rm max} = m_{\rm e}\gamma_{\rm max}$. The energy losses of electrons are dominated at the highest energies (close to $E_{\rm max}$) by the synchrotron process occurring in the magnetic field of the transition region.
The $\gamma$-ray production at these highest energies in the IC scattering of thermal radiation from the massive star occurs typically at a lower rate than the synchrotron process but at lower energies IC process can dominate. The IC $\gamma$-rays have large enough energies that in some cases they can initiate IC $e^\pm$ pair cascades in the radiation of the massive star. Such cascade can develop provided that the optical depths for electrons in the anisotropic radiation of the massive star are large enough. These optical depths have been calculated in the most general case of the anisotropic radiation of the massive star by e.g. Bednarek~(1997, 2000). They can be easily scaled for the case of the massive star with the arbitrary parameters (see Bednarek~2008). 

We developed the Monte Carlo code which is able to calculate
the synchrotron radiation from the primary electrons (magnetic field of the transition region), the secondary cascade $e^\pm$ pairs (in the magnetic field of the massive star), and the $\gamma$-ray spectra produced in IC $e^\pm$ pair cascade occurring in the anisotropic radiation of the massive star. This code base on the earlier code discussed for the binary systems (e.g. Bednarek~2000). In order to calculate the synchrotron spectrum produced by $e^\pm$ pairs inside the volume of the binary system, we have to define the distribution of the magnetic field around the massive star. The strength of the magnetic field as
a function of distance from the center of the massive star
can be described by the following equations (Usov \& Melrose, 1992),
\begin{eqnarray}
B(r) \approx B_\star \times \left\{ \begin {array}{ll}
(R_\star/r)^{3} , & R_\star \leq r < R_{\star,\rm A} ,  \\
R_\star^3/(R_{\star, \rm A} r^2) , & R_{\star, \rm A} < r , \end{array} \right.
\label{eq19}
\end{eqnarray}
\noindent
where $B_\star$ is the surface magnetic field of the massive star, $R_{\star, \rm A}$ is the Alfven radius within the massive star wind which can be derived by solving the equation,  $(1-R_\star/r_{\star, \rm A})=\xi 
(R_\star/r_{\star, \rm A})^4$. It has simple approximate solution, 
\begin{eqnarray}
R_{\star, \rm A} = R_\star \times \left\{ \begin {array}{ll}
1+\xi , & \xi \ll 1 \\
{\xi}^{1/4}, & \xi \gg 1, \end{array} \right.
\label{eq20}
\end{eqnarray}
\noindent
where $\xi = {B_\star}^2R_\star/(\dot{M} v_{\infty})$, $\dot{M}$ is the mass loss rate, and $v_{\infty}$ is the stellar wind velocity.
This approximation for the magnetic field structure is satisfactory inside the most important part of the binary system (i.e. within $\sim 10R_\star$) for the stars which 
typical rotation velocity is lower than $\sim 10\%$ of their wind velocity. 
Note that the larger values of the surface magnetic field of the massive star do not necessary mean larger values of the magnetic field at some distance from the stellar surface (as would be expected from the simple analysis of Eq.~\ref{eq19}) since B(r) depends also on the Alfven radius which in turn depends on the surface magnetic field and the mass loss rate of the massive star.

\section{The case of LSI +61 303}

As an example, we apply the model for the massive binary system LSI +61 303 which has been discovered as a TeV $\gamma$-ray source by the MAGIC Collaboration 
(Albert et al.~2006) and confirmed by the VERITAS Collaboration (Acciari et al.~2008). The TeV $\gamma$-ray emission from this binary system is observed only at
a specific range of phases close to the apoastron. No any TeV $\gamma$-ray emission
(above $\sim 200$ GeV) has been observed up to now close to the periastron passage. The multi-wavelength observations of LSI +61 303 show a hint of correlation between X-ray and TeV $\gamma$-ray emission (Albert et al.~2008).
LSI +61 303 belongs to the rare class of sources observed in the whole high energy range from X-rays up to TeV $\gamma$-rays (Tavani et al.~1996, van Dijk et al.~1996, Kniffen et al.~1997).

The massive star in this binary system belongs to a class B0 V. It is characterized by the  surface temperature $T_\star = 29850$ K and the radius $R_\star = 6.7 R_\odot$ (see e.g. Harmanec 1988). The basic parameters of the binary system are not exactly known. Two sets of parameters have been proposed by Casares et al.~(2005) and Grundstrom et al.~(2007). The semimajor axis, $a$, the eccentricity, $e$, the angle of periastron passage, $\omega_p$, the phase of the observer, $\omega_{\rm obs}$, the inclination angle, $\alpha$, the distance at periastron and apoastron, $r_{\rm p}$ and $r_{\rm a}$, are shown in Table 1.

\begin{quote}
\begin{table*}
  \caption{Basic parameters of the models.}
  \begin{tabular}{llllllllll}
\hline 
    &  $a\sin i$      & $\omega_p$ & $e$   & $\omega_{obs}$ & $\alpha$     
    & $r_{\rm p}$  & $r_{\rm a}$  & $\dot{M}_\star$ & $v_{\rm w}$\\
model I (Grundstrom et al.~2007)   &  8.4 $R_\odot$  & 0.30       & 0.55  & $33^{\rm o}$ 
& $25^{\rm o}$ & $1.33 R_\star$  & $4.60 R_\star$   & $3\times 10^{-8}M_\odot$ yr$^{-1}$ &
$10^3$ km s$^{-1}$ \\
model II (Casares et al.~2005)   &  8.2 $R_\odot$  & 0.23       & 0.72  & $70^{\rm o}$ 
& $15^{\rm o}$ & $1.32 R_\star$  &  $8.13 R_\star$  &  $10^{-8} M_\odot$ yr$^{-1}$ & $10^3$ km s$^{-1}$\\
\hline
\end{tabular}
  \label{tab1}
\end{table*}
\end{quote}

We perform the calculations of the synchrotron X-ray and the cascade IC $\gamma$-ray spectra for the location of the magnetar at the periastron and the apoastron passages for both sets of parameters. The surface magnetic field of the massive star is investigated in the range $B_\star = 10^{2-4}$ G and  the surface magnetic field of the magnetar is fixed to $B_{\rm NS} = 3\times 10^{14}$ G. Electrons are accelerated in the transition region with the differential power law spectrum extending to the maximum energies given by Eq.~\ref{eq14}.  For the illustration we chose the spectral index equal to -2 and the acceleration parameter $\xi_{-1} = 1$.
The power in relativistic electrons is estimated from Eq.~\ref{eq15}, applying specific  energy conversion efficiency $\eta$. Note, that this power scales with the product of $\eta P_1^{-2}$ which is taken in our calculations to be equal to $10^{-2}$.
We apply the inclination angles of the binary system to be consistent with other basic parameters (see Table 1) and assume the mass loss rate of the massive star equal to $3\times 10^{-8}M_\odot$ yr$^{-1}$ (model I) and $10^{-8}M_\odot$ yr$^{-1}$ (model II).
The results of the example calculations with the parameters mentioned above are shown in Fig.~1a-f. For all considered model parameters, the $\gamma$-ray spectra change drastically between the periastron and apoastron passages. No any emission above $\sim$1 TeV is observed at the periastron due to the inefficient acceleration of electrons.
The TeV $\gamma$-ray emission clearly extends through the TeV energy range at the apoastron, in agreement with the MAGIC and VERITAS observations. Note that $\gamma$-ray spectra depend also on the surface magnetic field of the massive star. For larger $B_\star$, secondary $e^\pm$ pairs from the IC cascade lose more energy on the synchrotron process. That's way, the synchrotron  X-ray spectra are on the higher level for larger $B_\star$.

\begin{figure*}
\vskip 12.2truecm
\includegraphics{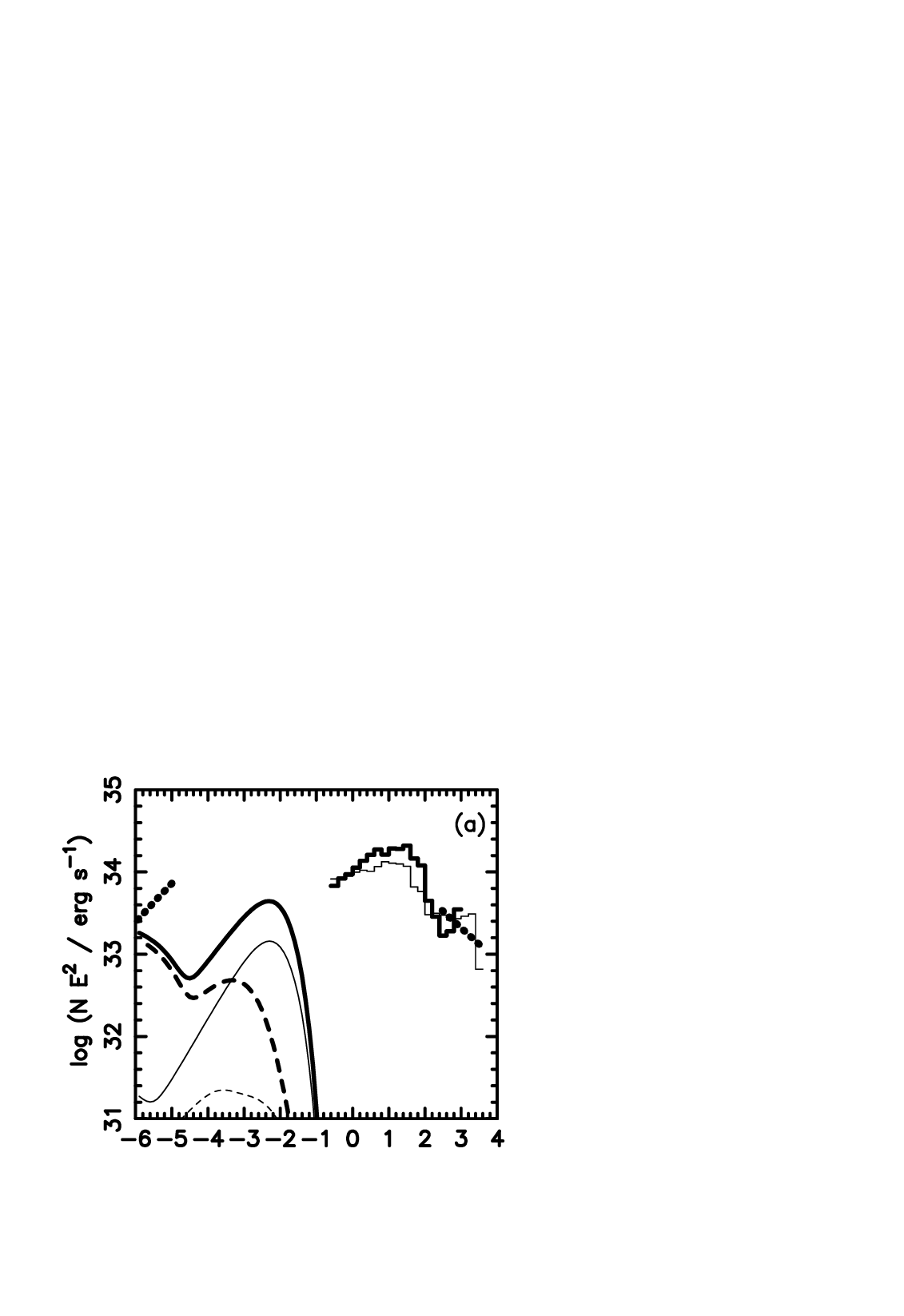}
\includegraphics{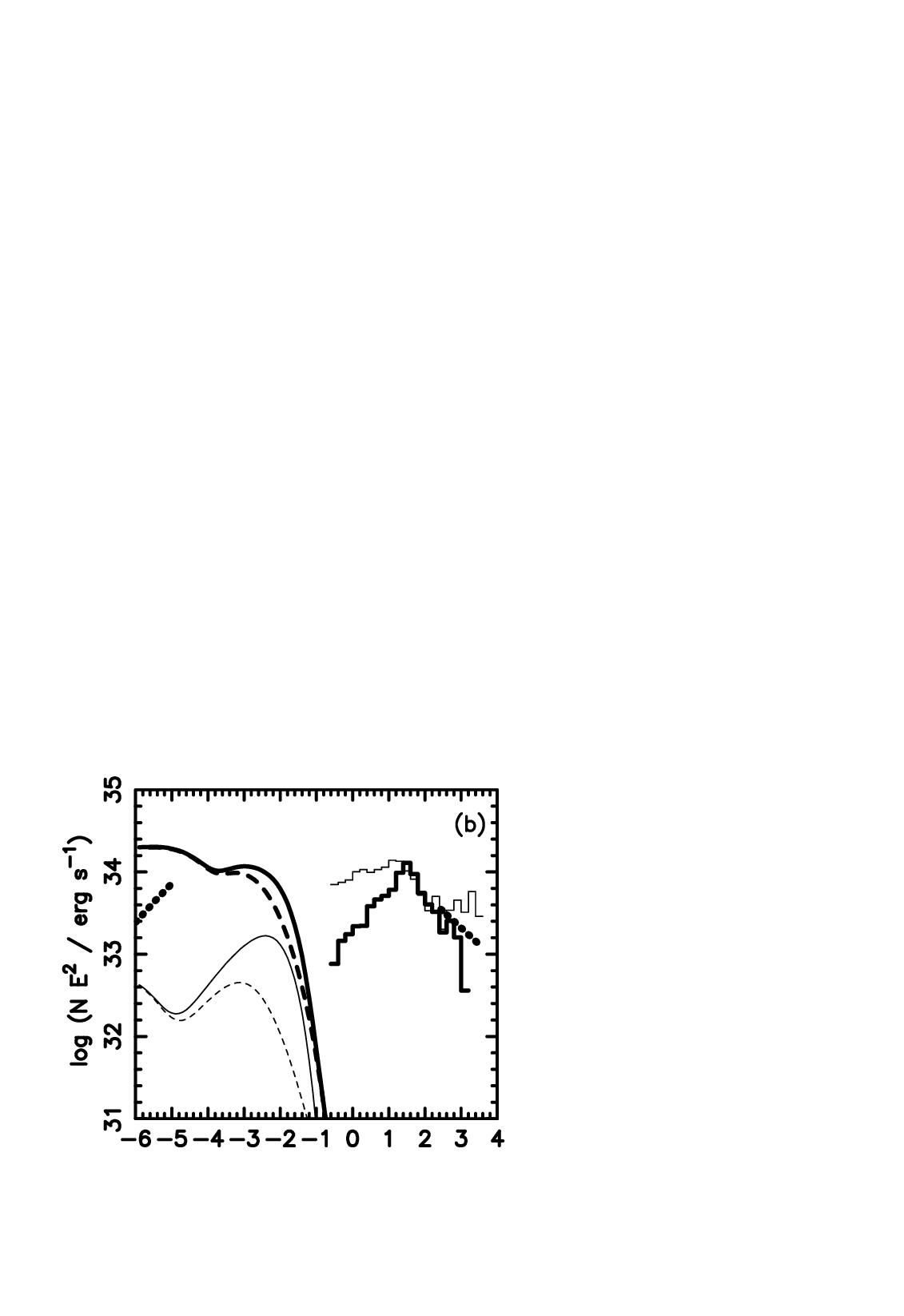}
\includegraphics{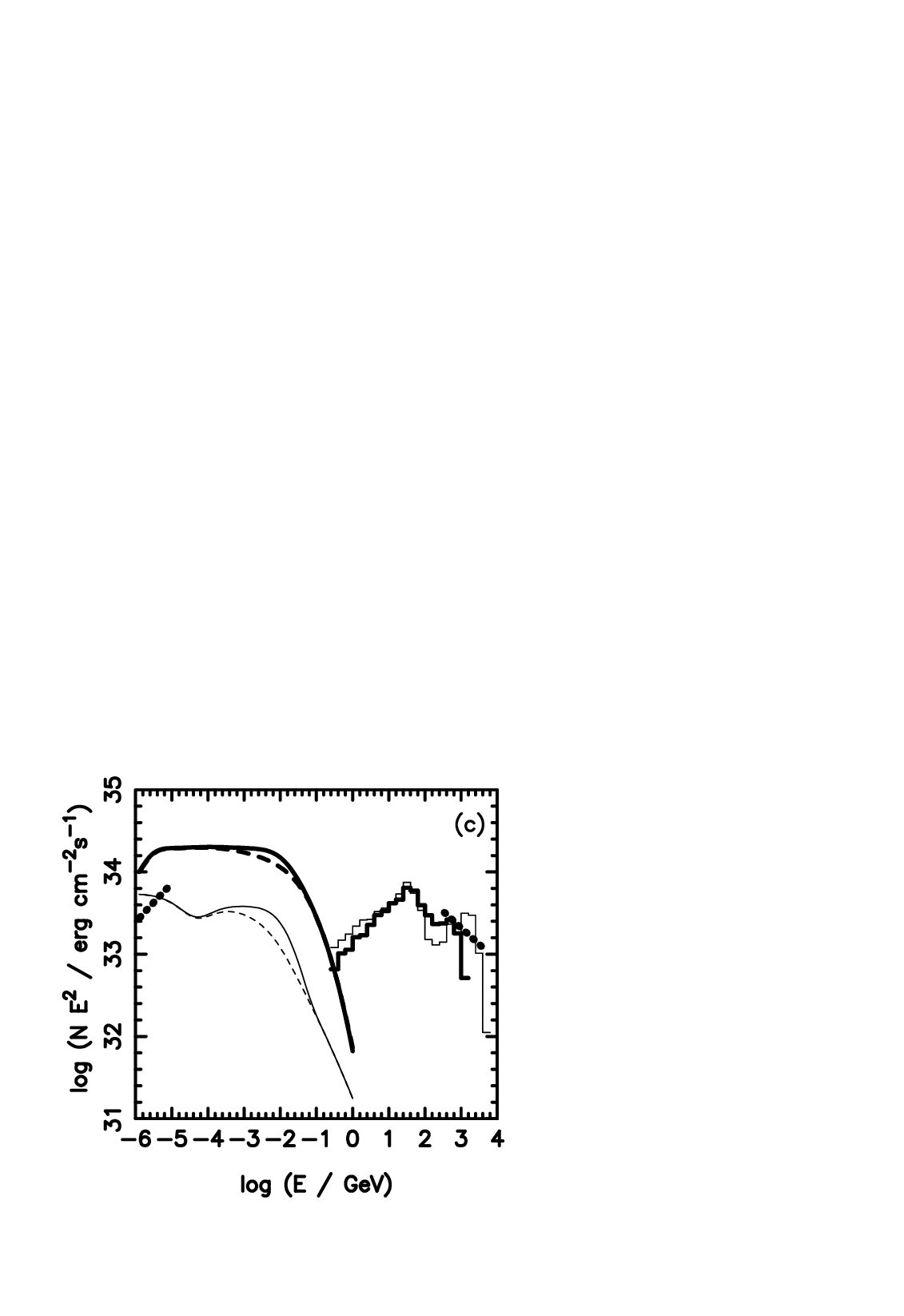}
\includegraphics{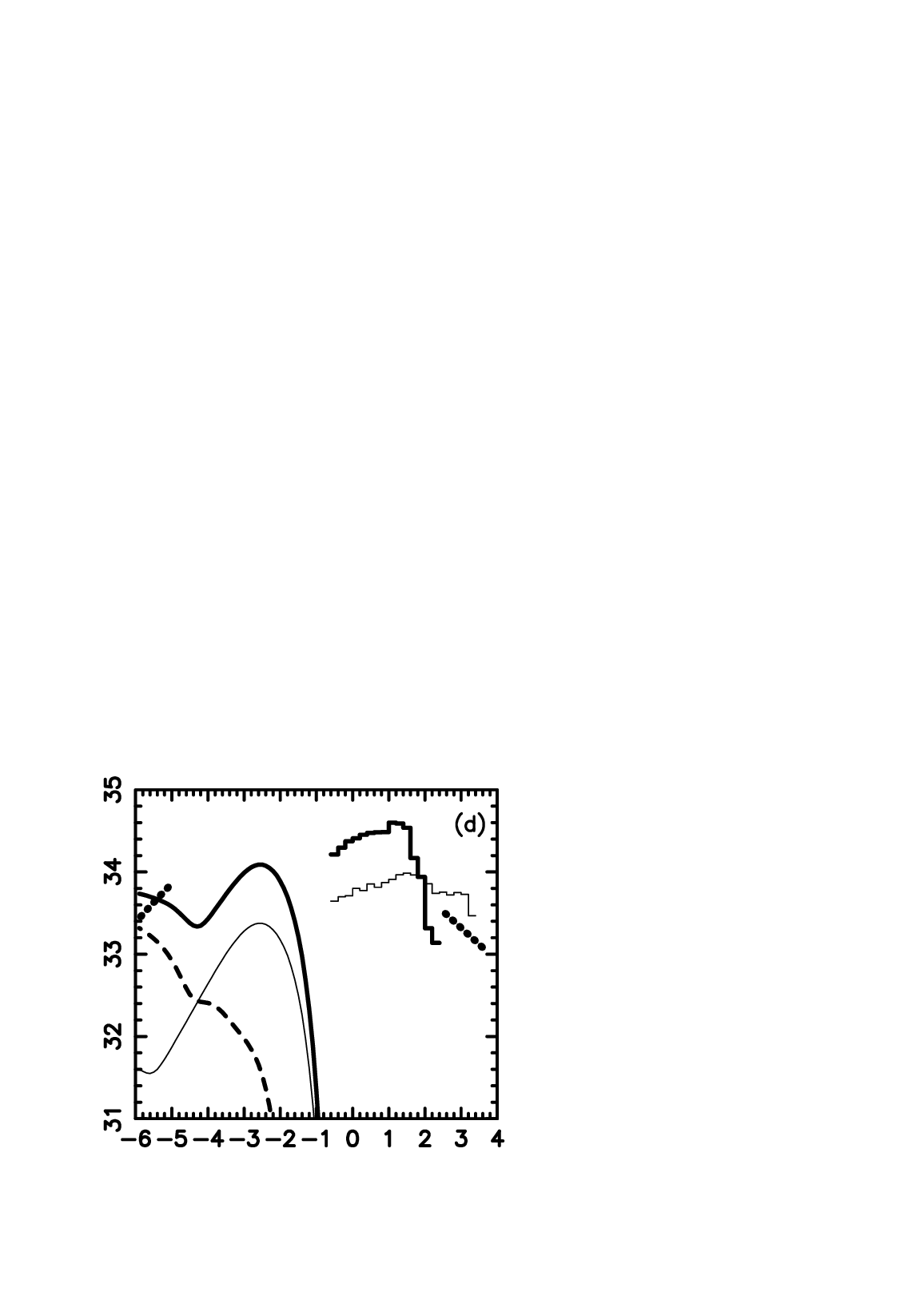}
\includegraphics{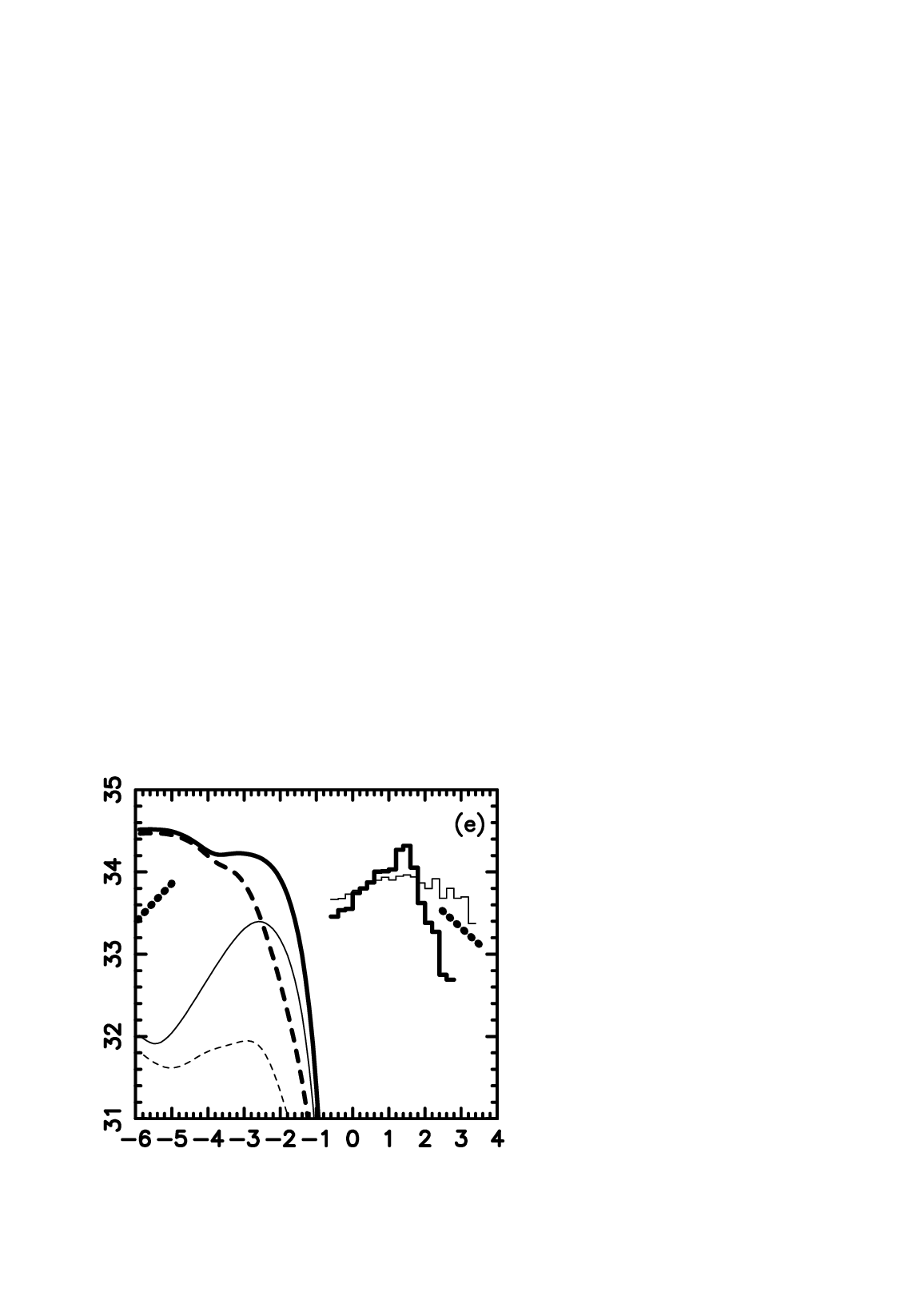}
\includegraphics{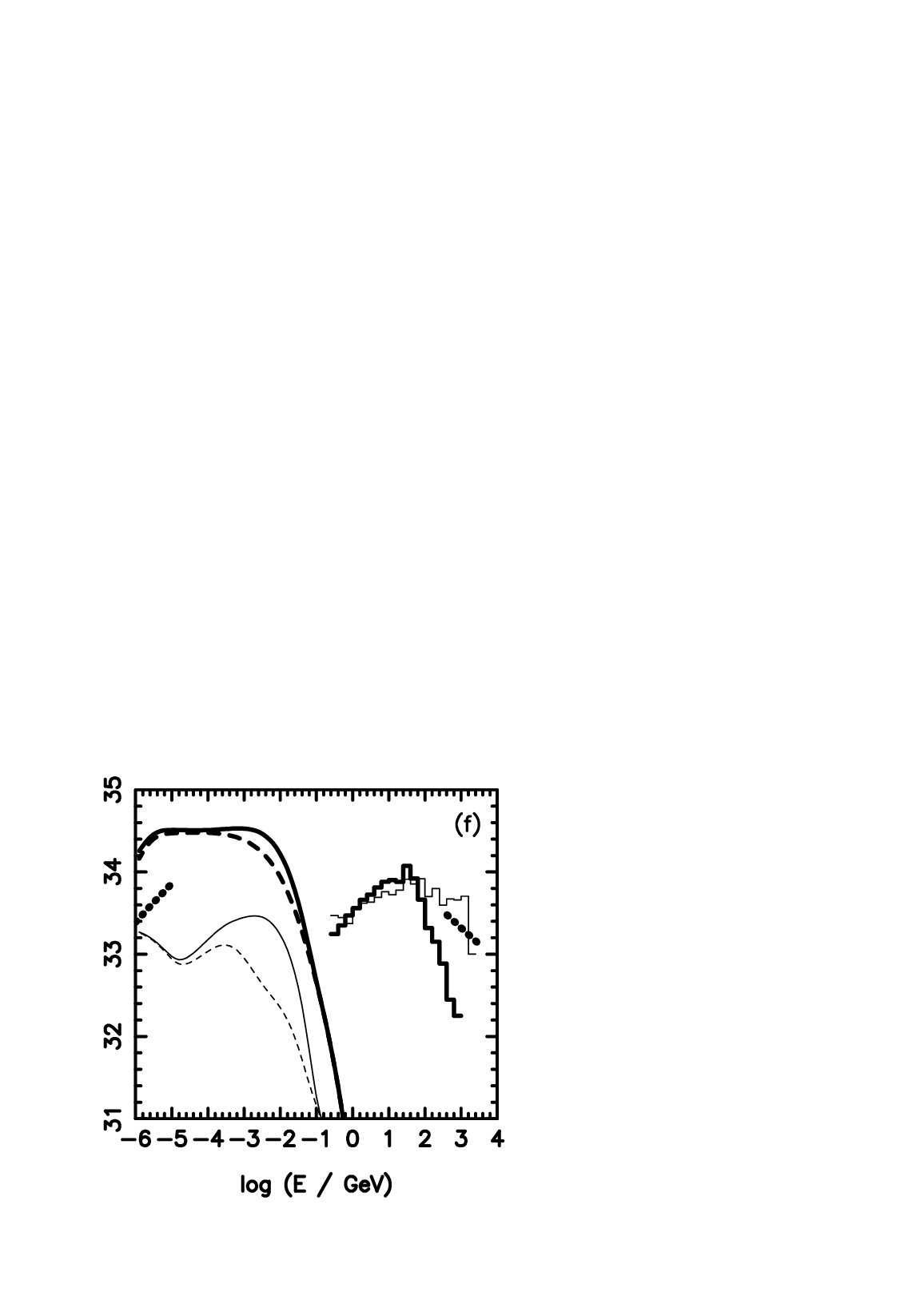}
\caption{The X-ray spectra (curves) and TeV $\gamma$-ray spectra (histograms) calculated for the periastron (thick) and apoastron (thin) passages
of the magnetar inside binary system LSI +61 303 for two different sets of the 
parameters (see Table 1): figures (a), (b), and (c) for the model I and (d), (e), and (f) for the model II. The surface magnetic field of the massive star is $B_\star = 10^2$ G (a) and (d); $B_\star = 10^3$ G (b) and (e), $10^4$ G (c) and (f).
The calculations have been done for $\eta p_1^{-2} = 10^{-2}$ and $\xi_{-1} = 1$.
For comparison we show the level of the X-ray and TeV $\gamma$-ray emission reported by RXTE (Smith et al.~2008) and MAGIC (Albert et al.~2006) and VERITAS (Acciari et al.~2008)   telescopes.}
\label{fig1}
\end{figure*}

The distances of the periastron passage, $r_{\rm p}$, for both models are very similar (Table 1), but distances of the apoastron passage, $r_{\rm a}$, differ significantly. This can explain the stronger absorption features in the case of the IC 
cascade spectrum at the apoastron in the case of parameters proposed by Grundstrom et al.~(2007). However the steeper cascade IC spectra obtained at the periastron for the Casares et al.~(2005) parameters has to be related to the differences in the angle of the periastron passage in respect to the location of the observer. Note, that in most of the considered cases (e.g. figures a,c,e,f), the GeV $\gamma$-ray fluxes does not change drastically between the periastron and apoastron passages. Therefore, it is likely that observed strong modulation of the TeV $\gamma$-ray emission with the period of the binary system may not be observed by the {\it Fermi} LAT detector in the GeV energies. 

The synchrotron spectra have more complicated shape due two the components produced by primary electrons and secondary $e^\pm$ pairs. The strength of the component produced by secondary $e^\pm$ pairs
strongly depends on the surface magnetic field of the massive star, $B_\star$, and on the distribution of the magnetic field around the star. In general, for stronger $B_\star$, the synchrotron spectrum from secondary $e^\pm$ pairs is stronger (see dashed curves in Fig.~\ref{fig1}), reaching saturation for $B_\star \sim 10^3$ G.
At the apoastron passage, the second synchrotron bump becomes comparable to the first bump only for the largest considered values of the surface magnetic field of the massive star,
i.e. $B_\star = 10^4$ G. 

Note that a part of primary electrons can likely escape from the transition region with the matter which can be expelled along the rotational axis
of the magnetar, i.e. within the possible transient jet which is sometimes expected in the case of the accretion process occurring in the propeller regime. The synchrotron emission of these electrons in the jet is not taken into account in this calculations.

In this simple model, the conversion coefficient of energy from the transition region to relativistic electrons, $\eta$, is kept independent on the phase of the binary system. However, $\eta$ may also depend on the accretion rate (i.e. a distance between the stars) in the way which is at present unknown. Therefore, any absolute normalization of the spectra is problematic since it
may vary with the phase of the binary system, e.g. as observed in the hard X-rays (Chernyakova et al.~2006). Only detailed phase dependent multi-wavelength observations can put constraints on the injection details of electrons at different phases of the binary system.

\section{Conclusion} 

We show that strongly magnetized, accreting neutron stars (magnetars) close to the massive stars 
inside the binary systems can be responsible for the acceleration of electrons up to TeV energies. These electrons interact with the magnetic field of the inner magnetosphere of the magnetar and with the soft radiation field from the massive companion producing X-rays (synchrotron process) and $\gamma$-rays (inverse Compton process). A part of injected $\gamma$-rays can be absorbed in the radiation of massive star developing IC $e^\pm$ pair cascade.
Secondary $e^\pm$ pairs lose significant part of their energy on the synchrotron emission. 
The $\gamma$-ray spectra produced in such scenario can extend up to TeV energies  and the synchrotron spectra can extend up to MeV energies. 
The processes discussed above concern to a specific stage in the evolution of the binary system,
when the accretion rate onto the neutron star is at the range which guarantee the propeller phase.
In fact, this phase may be short lived due to the variable wind from the massive star.
When the accretion rate is different, the magnetar can move to the ejector or the accretor phase during which different scenarios should be considered. For example in the accretor phase the magnetar can gain angular momentum with the accreting matter and  decrease its rotational period as expected in the case of the millisecond pulsars. 
We expect (based on the comparison of Eq.~\ref{eq16} and~\ref{eq17}) that the propeller phase of the accretion onto the neutron star, in which observable fluxes of $\gamma$-rays can be detected, is relatively short lived. This may explain  a small number of massive binaries discovered up to now in the TeV $\gamma$-rays. 
Applying the parameters of the binary system  LSI +61 303, 
we show that such a model can explain the TeV $\gamma$-ray emission observed close to the apoastron passage of the neutron star. The lack of observable TeV emission from the parts of the orbit close to the periastron is due to the inefficient acceleration of electrons and the suppression of the TeV $\gamma$-rays in the IC $e^\pm$ cascade process in which also 
the synchrotron energy losses play an important role.

\section*{Acknowledgments}
This work is supported by the Polish MNiSzW grant N N203 390834.


\label{lastpage}
\end{document}